\documentclass[preprint,12pt,a4paper]{elsarticle}

\usepackage{amssymb}
\usepackage{hyperref}
\usepackage{fullpage}
\usepackage[frozencache=true]{minted}
\usepackage{listings}
\usepackage{color}
\usepackage[font=footnotesize]{caption}
\newminted{java}{fontsize=\footnotesize,frame=lines,framesep=10pt,escapeinside=||}
\newminted{kotlin}{fontsize=\footnotesize,frame=lines,framesep=10pt,escapeinside=||}
\newcommand{\mi}[1]{\texttt{#1}}

\journal{SoftwareX}

\begin{document}

\begin{frontmatter}



\title{JISA: A Polymorphic Test-and-Measurement Automation Library}


\author[a]{William A. Wood}
\author[b]{Thomas Marsh}
\author[c]{Henning Sirringhaus}
\address[a]{waw31@cam.ac.uk}
\address[b]{tm709@cam.ac.uk}

\address{Cavendish Laboratory, University of Cambridge, Cambridge, CB3 0HE}

\begin{abstract}
JISA is a software library, written in Java, aimed at providing an easy,
flexible and standardised means of creating experimental control software for
physical sciences researchers. Specifically, with an emphasis on enabling
measurement code to be written in an instrument-agnostic way, allowing such
routines to be reused across multiple different setups without requiring
modification. Additionally, it provides a simple means of recording and handling
data, as well as pre-built graphical user interface (GUI) ``blocks'' to enable
the relatively easy creation of graphical control systems. Together these allow
users to quickly piece together test-and-measurement programs with coherent user
interfaces, without requiring much experience of such things.
\end{abstract}


\end{frontmatter}

\begin{table}[H]
    \centering
    \begin{tabular}{|l|p{6.5cm}|p{6.5cm}|}
    \hline
    \textbf{Nr.} & \textbf{Code metadata description} &  \\\hline
    C1 & Current code version &  1.0\\\hline
    C2 & Permanent link to code/repository used for this code version &  {https://github.com/OE-FET/JISA} \\\hline
    C3 & Code capsule & N/A \\\hline
    C4 & Legal Code License   & AGPLv3 \\\hline
    C5 & Code versioning system used & git \\\hline
    C6 & Software code languages, tools, and services used & Java\\\hline
    C7 & Compilation requirements, operating environments and dependencies & JDK 11 or newer\\\hline
    C8 & Developer documentation/manual & {https://github.com/OE-FET/JISA/wiki} \\\hline
    C9 & Support email for questions & waw31@cam.ac.uk\\\hline 
    \end{tabular}
    \caption{Code metadata.}
    \label{} 
    \end{table}

\raggedbottom

\section{Introduction and Motivation}

Writing software to control experimental systems, where multiple, individual
instruments are made to work in concert to perform automated measurement
routines, is commonplace in physical sciences research. When considering how
such software controls each instrument, one can split the process into three
parts. The first of these is the instrument itself, which listens to
instructions sent to it from the controlling computer --- often responding by
taking some action and/or replying \cite{IEEE488,IEEE4882,SCPI}. The second is
the ``driver'', the software library used to format and send these commands over
some hardware protocol and interpret responses from the instrument. The main
purpose of this is to present the functionality of the instrument, as well as
any output it may give, in a way that is readily useable by a programmer. The
third part we shall label the ``routine'', where the logic of a measurement is
laid out. This is essentially the software that makes use of the driver to
control the instrument as part of some wider measurement routine. Boundaries
between entities (such as those listed here) and how information flows through
them, are referred to as ``interfaces'' in computer
science \cite{IEEEDictionary}. Therefore, there are two interfaces in this
picture: Instrument\,$\leftrightarrow$\,Driver (ID), and
Driver\,$\leftrightarrow$\,Routine (DR).

For instruments, their ID interfaces are defined by their manufacturer, almost
always with some degree of arbitrariness. As a result, two instruments with the
same functionality, but of different make and model, are unlikely to have
compatible ID interfaces, and thus will require different drivers. Furthermore,
because drivers, and thus DR interfaces, are normally designed to follow the
same rough structure as the ID interface of their target instrument, the same
incompatibility often applies to DR interfaces too. The result of this is that a
routine written to perform an experiment using one set of instruments would need
to be significantly altered to accommodate instruments of the same type, but
different makes and models, thus often preventing measurement code from being
reusable. 

\begin{figure*}[htb]
    \centering
    \includegraphics[width=\linewidth]{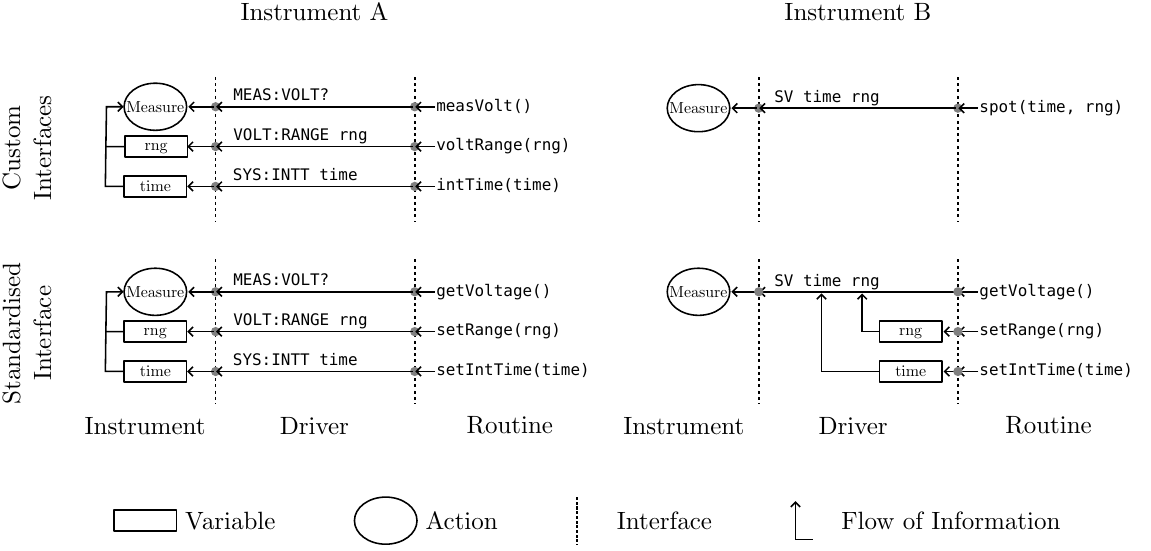}
    \caption{Diagrammatic representations of the ID and DR interfaces for two instruments, A and B, where the methods exposed by the driver to the measurement routine are represented as information sent from the routine to the driver over the DR interfaces, where they are converted into the specific command codes required by the instrument, and then sent to the instrument over the ID interface. The top row represents a typical DR interface made to simply mimic the ID interface, whereas the bottom row shows how both instruments could be abstracted behind a common interface allowing interchangeability.}
    \label{fig:interfaces}
\end{figure*}

As an illustrative example, let us suppose we have two voltmeters: A and B.
Voltmeter A holds the settings of measurement range, ``\mi{rng}'', and
integration time, ``\mi{time}'', persistently in its own memory, which it
recalls whenever it is instructed to take a measurement. Voltmeter B, on the
other hand, treats the task more like a one-off spot measurement, with both
parameters needing to be specified in the measurement command each time. If
drivers were to be written for these instruments, with their DR interfaces made
to mimic the ID interfaces defined above, then they will look drastically
different. The driver for A will have three methods, two for individually
setting the \mi{rng} and \mi{time} settings, and one for triggering a
measurement, whereas B will only have one method which takes both \mi{rng} and
\mi{time} as arguments. This is depicted in the top row of
Figure~\ref{fig:interfaces}.

Clearly, any measurement routine using the driver for A would require
significant alteration to instead accommodate B. However, this can be avoided by
defining a common interface, and hiding the differences in the underlying
implementation behind it. For instance, the \mi{rng} and \mi{time} settings for
B could be held as variables within the driver layer, with them being recalled
and sent any time a voltage measurement is requested (as well as holding initial
default values). That way, both A and B can implement a common DR interface, as
is shown on the bottom row of Figure~\ref{fig:interfaces}. Now, a measurement
routine can be written that will work regardless of whether it is using A or B
as, from the perspective of the routine, they are now indistinguishable. In
essence, we have taken the concept of a voltmeter, and turned it into a black
box by defining a common voltmeter interface. In rough terms, this is what is
known as ``polymorphism'' in object-oriented programming: where one has
multiple, interchangeable objects of different classes that all implement the
same interface.

Most available tools and drivers tend to lack any such standardisation. For
instance, in National Instruments' LabVIEW, instrument drivers are typically
written in a completely procedural manner, using no object-oriented features at
all. On the other hand, open-source drivers (normally written in Python), while
typically object-oriented, do not use any common interfaces. There are some
projects that have tried to standardise such things, namely PyMeasure for Python
\cite{PyMeasure}, but this has not been their primary focus. Furthermore, Python
has long lacked static typing features, making it a less than ideal language for
defining such structures and enforcing them on later additions to the codebase
(although this has been changing in more recent years).

In contrast, JISA aims to standardise from the ground up. It does this by
defining standard interfaces to define the functionality of different ``types''
of instrument (i.e. voltmeters, temperature controllers, etc) as well as a
growing set of drivers that implement them. For instance, if someone wants to
write a driver for a voltmeter in JISA, it is required to meet the standard
voltmeter specification, ensuring compatibility between different makes and
models.

Additionally, JISA provides basic data-recording and data-handling
functionality, as well as a simplified means of creating graphical user
interfaces (GUI) for test-and-measurement programs. This comes in the form of a
series of pre-defined ``blocks'' that can be easily connected together to make a
cohesive user interface. The idea being to provide a means for other scientists
to make control programs quickly and easily, without requiring extensive
knowledge or skill in the subject.

\section{Software Description}

JISA is written in Java, thus it is useable in any language that can be
interpreted on, ``hook into'', or be compiled to be run directly on the Java
Virtual Machine (JVM). Examples include Kotlin (compiles to Java bytecode), and
Python (can either compile through Jython/GraalVM, or hook in through JPype).
This makes it interoperable with almost any language, especially since the
development of the GraalVM polyglot JVM, which is able to compile and run code
written in many different languages, giving them all access in one way or
another to Java classes and libraries. This includes: Python, JavaScript, Ruby,
PHP, R, and C/C++ (via LLVM). Java is also a well-known programming language
itself, often being the language chosen for teaching object-oriented
programming, both in schools and university STEM courses \cite{Mahmoud2004}.
However, example code listings in this paper shall be written in Kotlin, to keep
them compact. In general, JISA comprises three parts: standardised instrument
control, basic data handling and processing, and GUI creation. In this section
we shall explore each of these.

\subsection{Instrument Control}

In JISA, each physical instrument is represented by an object, instantiated from
a class representing its make and model. This way, instrument objects can take
advantage of the polymorphism enabled by object orientation. For instance, in
Java and most other programming languages, classes can be defined as
``implementing'' an ``interface''. In this sense of the word, an interface is
like a template, specifying what methods any classes that implement it must
themselves implement. That is, it specifies (at least part of) the interface, in
the previous sense of the word, of objects instantiated from the class, hence
the name. Therefore, instrument objects in JISA implement interfaces that
describe what type of instrument it is. For instance, a class for controlling a
source-measure unit (SMU) would implement the \mi{SMU} interface, which itself
is defined by combining the interfaces of voltmeters, ammeters, voltage sources
and current sources. Therefore, any SMU that is controlled in JISA can be
thought of as being an object of type \mi{SMU}, \mi{VMeter}, \mi{IMeter},
\mi{VSource}, and/or \mi{ISource}, without any thought having to be given as to
which make and model it is. This enables one to write measurement routines in an
instrument-agnostic way, where the routine itself simply expects to be handed an
instrument of a given type, not specific make/model. An example of this is
demonstrated in Listing~\ref{lst:conductivity}, where a conductivity sweep
routine expects an object of type \mi{SMU} to be passed as one of its arguments,
meaning that any make/model of SMU can be used.

\begin{listing*}[h!]

\begin{kotlincode}
fun sweep(smu: SMU, vals: Iterable<Double>, del: Long, table: ResultTable) {
    
    // Find columns in table with matching title and data type (explained later)
    val I = table.findColumn("Current", Double::class)
    val V = table.findColumn("Voltage", Double::class)

    smu.setFourProbeEnabled(false)     // Two-wire measurement
    smu.setIntegrationTime(20e-3)      // 20ms int time (or nearest available)
    smu.setCurrent(0.0)                // Source current at 0 A
    smu.useAutoRanges()                // Auto range for both current and voltage
    smu.turnOn()                       // Enable SMU output

    for (current in vals) {            // Loop over each current value

        smu.setCurrent(current)        // Source current sweep value
        Thread.sleep(del)              // Wait for delay time

        table.addRow { row ->
            row[I] = smu.getCurrent()  // Ask SMU for current measurement
            row[V] = smu.getVoltage()  // Ask SMU for voltage measurement
        }
    }

    smu.turnOff()                      // Disable SMU output

}

val smu1 = K2450(...)                         // Keithley 2450 SMU
val smu2 = Agilent4155B(...).getSMUChannel(0) // First SMU in an Agilent SPA

// Run the same measurement routine using two different SMUs without changes
sweep(smu1, Range.linear(0.0, 1e-6, 5), 50, table)
sweep(smu2, Range.linear(0.0, 1e-6, 5), 50, table)
\end{kotlincode}

\caption{A measurement routine that expects to be given as arguments: (1)
\mi{smu}: an object representing the SMU to use, (2) \mi{vals}: an iterable
(i.e. can be looped over) collection of values to sweep current through, (3)
\mi{del}: the delay time to use between setting the current and measuring the
voltage in ms, and (4) \mi{table}: a pre-defined data structure to store the
results of the sweep in.\label{lst:conductivity}}

\end{listing*}

Further to this, each instrument class extends from a base class, representing
the underlying connection to the instrument. This makes sure that each
instrument class inherits the communications methods it requires. For instance,
most instruments extend the \mi{VISADevice} base class, which enables
connections to be made via an underlying system VISA library, such as those
provided by National Instruments, Keysight, and Rohde \& Schwarz, or more
directly via native serial ports, linux-gpib, Java TCP-IP sockets, or libUSB if
no VISA library is available or otherwise able to make the connection
successfully.

When instantiating an instrument object, an address is specified to indicate how
the instrument is connected to the computer. These are in the form of address
objects, which all implement the \mi{Address} interface. Therefore, if one was
to connect to an instrument over, for instance, TCP-IP at address
``192.168.0.5'' and port number 5555, one would need to create a
\mi{TCPIPAddress} object and pass it those two parameters as constructor
arguments. This is shown in Listing~\ref{lst:addresses}, where multiple examples
of creating address objects are shown.

\begin{listing*}[h!]
\begin{kotlincode}
val a = TCPIPAddress("192.168.0.5", 5555)     // TCP-IP
val b = LXIAddress("192.168.0.6")             // LXI / VXI-11
val c = GPIBAddress(0, 24)                    // GPIB
val d = SerialAddress("/dev/ttyS0")           // Serial (Unix)
val e = SerialAddress("COM5")                 // Serial (Windows)
val f = SerialAddress(3)                      // Serial (ASRL)
val g = USBTMCAddress(0x0FE, 0x2600)          // USB-TMC
val h = USBRawAddress(0x0FE, 0x2600)          // USB Raw
val i = Address.parse("VISA::ADDRESS::INSTR") // Parse from VISA string
\end{kotlincode}
\caption{Various different address objects representing different physical
connections to instruments.\label{lst:addresses}}
\end{listing*}

To connect to an instrument, the constructor for the class matching its make and
model must be called, giving it the relevant address object to tell it where to
connect to. For instance, to connect to a Keithley 2450 SMU, connected over GPIB
(board 0, primary address 12), one would create a \mi{K2450} object by calling
its constructor, passing it a \mi{GPIBAddress} object, as is shown at the top of
Listing~\ref{lst:connect}.

\begin{listing*}[h!]
\begin{kotlincode}
val a = K2450(GPIBAddress(0, 12))              // Keithley 2450 SMU
val b = MercuryITC(LXIAddress("192.168.0.14")) // Mercury ITC
val c = K2182(SerialAddress("/dev/ttyUSB0"))   // Keithley 2182 nano-voltmeter
val d = LS336(Address.parse("ASRL5::INSTR"))   // LakeShore 336 TC
\end{kotlincode}
\caption{Connecting to different instruments over different
connection/communication protocols.\label{lst:connect}}
\end{listing*}

For instruments that are collections of sub-instruments (for instance, a
temperature controller is effectively a collection of PID loops, thermometers,
and heaters), the \mi{MultiInstrument} interface exists. This requires any
implementing class to provide methods that allow for these sub-instruments to be
queried and extracted. Specifically, they must have a method that returns a list
of all sub-instrument types/classes they contain, as well as a method that takes
an instrument type and returns a list of all sub-instruments that match that
type. This is demonstrated in Listing~\ref{lst:subinstrument}.

\begin{listing*}[h!]
\begin{minipage}[t]{0.49\linewidth}
\begin{kotlincode}
if (inst.contains(IMeter::class)) {       
    val i = inst.get(IMeter::class, 0);
}
\end{kotlincode}
\end{minipage}
\hfill
\begin{minipage}[t]{0.49\linewidth}
\begin{kotlincode}
if (IMeter::class in inst) {       
    val i = inst[IMeter::class, 0]
}
\end{kotlincode}
\end{minipage}
\caption{Example code where a connected instrument is being checked to see if it
contains any sub-instruments that could be used as an ammeter (\mi{IMeter}), if
so then the first one is chosen from a list of all such sub-instruments. Using
both ``Java style'' (left) and ``Kotlin style'' (right)
syntax.\label{lst:subinstrument}}
\end{listing*}

\subsection{Data Handling and Configurations}

The second part of the library is a simple means of recording, handling and
writing measurement data, as well as simple measurement control structures. The
data handling is implemented through the \mi{ResultTable} abstract class, which
is designed to represent a table of data. This has two implementations:
\mi{ResultList} (for holding data in memory), and \mi{ResultStream} (which uses
a backing file to store and retrieve data directly). Data added to such objects
can then easily be written as CSV files, either through a method call, or by
choosing the \mi{ResultStream} implementation of \mi{ResultTable}.

To create a \mi{ResultTable}, one must first define the columns it is to
contain. This is done by using the various \mi{Column.of...()} methods. Which
one to use is chosen depending on the data type the column is meant to hold.
These can then be supplied as constructor parameters to either \mi{ResultList}
or \mi{ResultStream} to build a table of these columns that is ready to accept
data. This is all shown in Listing~\ref{lst:tables}.

\begin{listing*}[h!]

\begin{kotlincode}
val C = Column.ofIntegers("Count")        // Integers
val V = Column.ofDoubles("Voltage", "V")  // Decimals (Doubles)
      = Column.ofDecimals("Voltage", "V") // alias for ofDoubles()
val N = Column.ofStrings("Notes")         // Strings (Text)
      = Column.ofText("Notes")            // alias for ofStrings()
val B = Column.ofBooleans("Switch")       // Booleans: true/false

val list   = ResultList(C, V, N, B)
val stream = ResultStream("/path/to/file.csv", C, V, N, B)
\end{kotlincode}
\caption{The four main column types supported by \mi{ResultTable} objects
(integer, decimal, text, and boolean) which are then given to the two
implementations of \mi{ResultTable} (\mi{ResultList} and \mi{ResultStream}) to
construct tables using said columns.\label{lst:tables}}
\end{listing*}

To add data to such structures, there are a few options. One can either use
\mi{addData(...)} to add rows of data in column order, or use \mi{addRow(...)}
or \mi{mapRow(...)} to add them in any order by use of lambda expressions or
maps respectively, shown in Listing~\ref{lst:tableadd}.

Data stored in a \mi{ResultTable} can then be accessed, either in order by
iterating over each row, or by random access by using its accessor/get method.
Each row is represented by a \mi{Row} object, within which each column value can
be accessed by supplying its column object as an index. This is demonstrated in
Listing~\ref{lst:tableaccess}.

Finally, \mi{ResultTable} objects can have extra info assigned to them as
``attributes'', be output to CSV files, and loaded back in from said files. This
is either done by calling \mi{output(...)} on the object, or by using the
\mi{ResultStream} implementation. Loading back in is achieved by using static
\mi{loadFile(...)} methods. When loading in a \mi{ResultTable}, we often need to
extract the relevant column reference objects from it. For this purpose, there
are the \mi{findColumn(...)} methods. This is shown in
Listing~\ref{lst:tablewrite}.

\begin{listing*}[h!]
\begin{minipage}[t]{0.49\linewidth}
\begin{kotlincode}
list.addData(1, 12.0, "Text", true)

list.addRow { row ->
    row[B] = true   
    row[C] = 1      
    row[N] = "Text" 
    row[V] = 12.0   
}
\end{kotlincode}
\end{minipage}
\hfill
\begin{minipage}[t]{0.49\linewidth}
\begin{kotlincode}
list.addData(1, 12.0, "Text", true)

list.mapRow(
    B |\textcolor{blue}{\emph{to}}| true,
    C |\textcolor{blue}{\emph{to}}| 1,
    N |\textcolor{blue}{\emph{to}}| "Text",
    V |\textcolor{blue}{\emph{to}}| 12.0
)
\end{kotlincode}
\end{minipage}
\caption{Adding data to a \mi{ResultTable} by specifying it in column order
(first line in both), out of order by use of a lambda (bottom left), or by
mapping columns to values (bottom right), assuming table structures defined
previously in Listing~\ref{lst:tables}.\label{lst:tableadd}}
\end{listing*}

\begin{listing*}[h!]
\begin{minipage}[t]{0.49\linewidth}
\begin{kotlincode}
// Plain Java-style method call
val row = list.get(0)

// Kotlin accessor syntax
val row = list[0]

// Iterate over each row in order
for (row in list) {
    ...
}
\end{kotlincode}
\end{minipage}
\hfill
\begin{minipage}[t]{0.49\linewidth}
\begin{kotlincode}
// Plain Java-style method call
val voltage = row.get(V)

// Kotlin accessor syntax
val note = row[N]

// Iterate over each column in order
for ((column, value) in row.values) {
    ...
}
\end{kotlincode}
\end{minipage}
\caption{Accessing rows, columns and values stored within a \mi{ResultTable} by
using plain Java-style method calls, Kotlin accessor syntax, and by iterating
over all values in order, assuming table structures defined previously in
Listing~\ref{lst:tables}.\label{lst:tableaccess}}
\end{listing*}

\begin{listing*}[h!]
\begin{kotlincode}
list.setAttribute("Integration Time", "20 ms")     // Store extra info in table
list.setAttribute("Creator", "Bob")

list.output("/path/to/file.csv")                   // Write to CSV file
 
val l = ResultList.loadFile("/path/to/file.csv")   // Load from CSV file
val s = ResultStream.loadFile("/path/to/file.csv") // Stream CSV file

val V = l.findColumn("Voltage", Double::class)     // Extract column references
val N = l.findColumn("Notes", String::class)       // by specifying names and
val B = l.findColumn("Switch", Boolean::class)     // data types
val C = l.findColumn("Count", Int::class)

val time    = l.getAttribute("Integration Time")   // time    = "20 ms"
val creator = l.getAttribute("Creator")            // creator = "Bob"
\end{kotlincode}
\caption{Storing attributes in a \mi{ResultTable}, writing to CSV files, loading
from CSV files, extracting column references, and reading stored
attributes.\label{lst:tablewrite}}
\end{listing*}
\newpage
For measurement control structures, JISA provides a queue infrastructure,
allowing for individual measurement routines to be strung together into more
complex overall routines, with error handling built in. It also provides
configurable \mi{Connection}, \mi{Configuration}, and \mi{ConfigFile} structures
that allow for the storing of instrument connection settings, instrument
parameters, and general storing of other variables to a JSON (JavaScript Object
Notation) configuration file, allowing for persistent settings between runs of a
program.

\subsection{Graphical User Interfaces}

\begin{listing*}[p]

\begin{kotlincode}
fun main() {

    // Asks and waits for the user to connect to and configure an SMU
    val smu = GUI.askUserForInstrument("Please connect to an SMU", SMU::class)

    // Define columns for results table
    val I = Column.ofDecimals("Current", "A")
    val V = Column.ofDecimals("Voltage", "V")

    // Create table of results using columns
    val results = ResultList(I, V)

    // Create user input, table and plot, as well as a grid to hold them together
    val fields   = Fields("Parameters")
    val delay    = fields.addTimeField("Delay Time")
    val currents = fields.addDoubleRange("Currents [A]")
    val table    = Table("Table of Results", results)
    val plot     = Plot("Plot of Results", results)
    val grid     = Grid("Current Sweep", 3, fields, table, plot)

    // Add toolbar button to run sweep method when clicked, extracting the
    // values currently written in the input fields by calling .get()
    grid.addToolbarButton("Run") {
        sweep(smu, currents.get(), delay.get(), results)
    }

    // Show the grid as a window, terminate program if closed
    grid.setExitOnClose(true)
    grid.show()

}
\end{kotlincode}
\centering
\includegraphics[width=\linewidth]{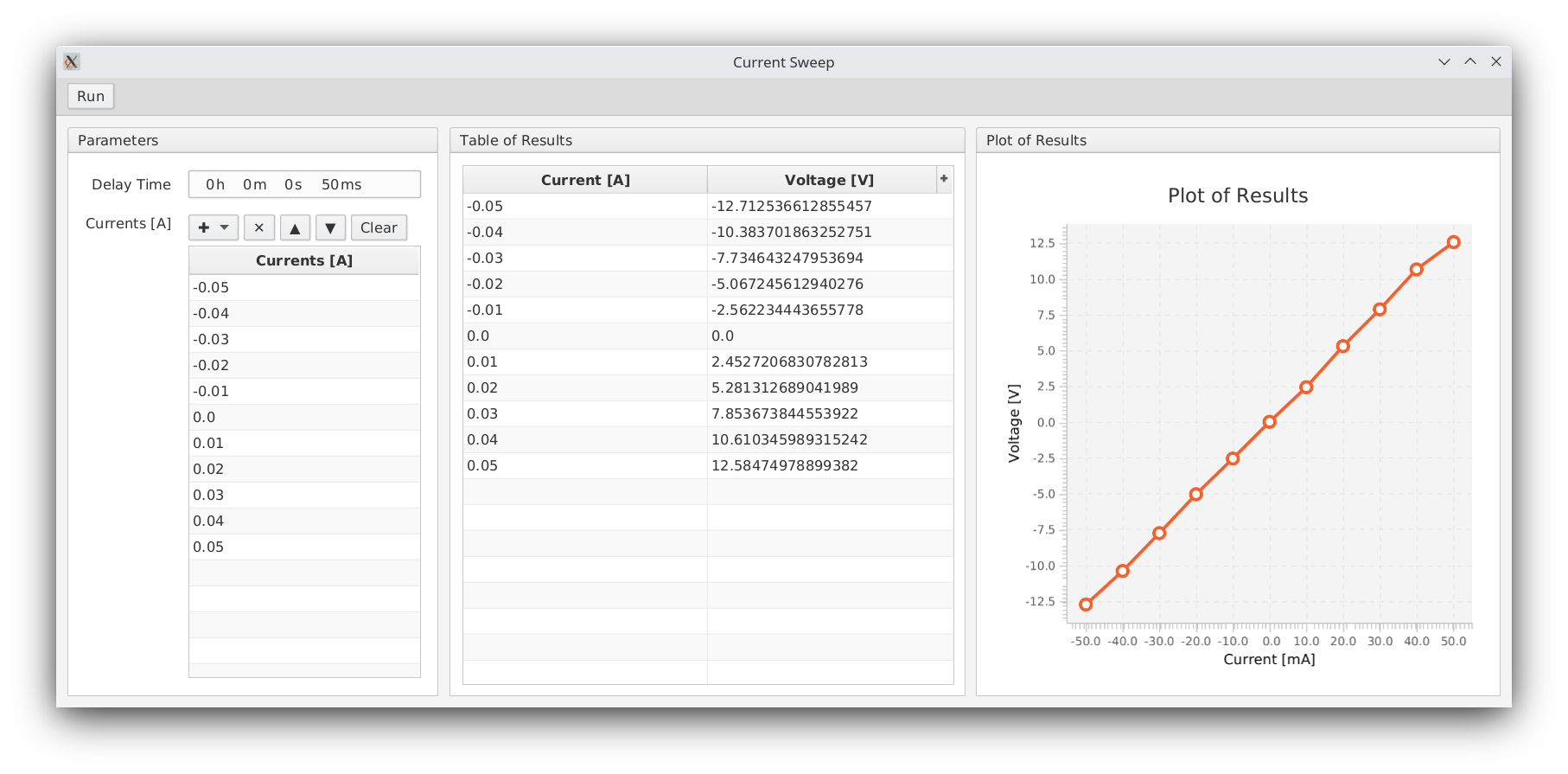}
\caption{An example Kotlin program to perform the previously defined current sweep while providing a GUI to control it (top) with screenshot of it running --- after ``Run'' is clicked (bottom).}
\label{lst:gui}
\end{listing*}

The third part of the library is that which allows GUIs to be built from a set
of pre-defined ``blocks'' or ``elements'', based on JavaFX. For instance, a user
can create a \mi{Plot} object (which uses the ChartFx library\cite{ChartFX}),
and tell it to watch a \mi{ResultTable}. The end result is a plot of the data in
the \mi{ResultTable} that updates live. Similar blocks exist for displaying
\mi{ResultTable} objects in tabular form (\mi{Table}), and acquiring user input
(\mi{Fields}). There are also GUI element classes named \mi{Connector} and
\mi{Configurator} which wrap around the previously mentioned \mi{Connection} and
\mi{Configuration} objects to provide a graphical means for a user to connect to
and configure an instrument.

Each GUI element can be shown in its own window by calling its \mi{.show()}
method, or grouped together within \mi{Container} elements which arrange
multiple elements together into a single element. For instance, a \mi{Grid}
object can be created which will display multiple other GUI elements together in
a gridded layout, or a \mi{Tabs} object for arranging them into separate tabs.
Containers are considered to be GUI elements themselves, and thus can be added
to each other to create more complex layouts. An example of a simple GUI for
controlling a current sweep measurement is shown in Listing~\ref{lst:gui}. When
the ``Run'' button in the toolbar is clicked, the measurement routine from
Listing~\ref{lst:conductivity} is run, populating the \mi{ResultTable} that the
\mi{Table} and \mi{Plot} elements are watching, causing them to update live.

Additionally, JISA provides some structures to make the development of larger
applications based on it somewhat easier. For instance, it provides the
\mi{Action}, \mi{Measurement}, and \mi{ActionQueue} classes. Together, these
provide a skeleton for creating systems where individual actions and measurement
routines are queued to run sequentially (as well as in loops to create sweeps),
allowing the user to define larger overall measurement routines. This is
complemented with the \mi{ActionQueueDisplay} and \mi{MeasurementConfigurator}
GUI classes, which together allow users to graphically queue and configure their
measurement routines. An example of using an \mi{ActionQueue} with an
\mi{ActionQueueDisplay} is shown in Listing~\ref{lst:queue}.

\begin{listing*}[p]
\begin{kotlincode}
val queue   = ActionQueue()
val display = ActionQueueDisplay("Queue", queue)
display.addToolbarButton("Stop") { queue.stop() }
display.show()

val tChange = SimpleAction("Change Temperature") {
    loop.setSetPoint(100.0)
    loop.waitForStableValue(100.0, 1.0, 600000)
}

val vChange = SimpleAction("Change Voltage") {
    smu.setVoltage(10.0)
    smu.turnOn()
}

val ivMeasure = SimpleAction("Measure Voltage and Current") {
    val current = Repeat.prepare(10, 100) { smu.getCurrent() }
    val voltage = Repeat.prepare(10, 100) { smu.getVoltage() }
    Repeat.runTogether(current, voltage)
}

queue.addActions(tChange, vChange, ivMeasure)
when (queue.start()) {
    COMPLETE    -> {/* Completed without error */}
    INTERRUPTED -> {/* Interrupted early by calling stop() */}
    ERROR       -> {/* Encountered error(s) */}
}
\end{kotlincode}
\centering
\includegraphics[width=0.65\linewidth]{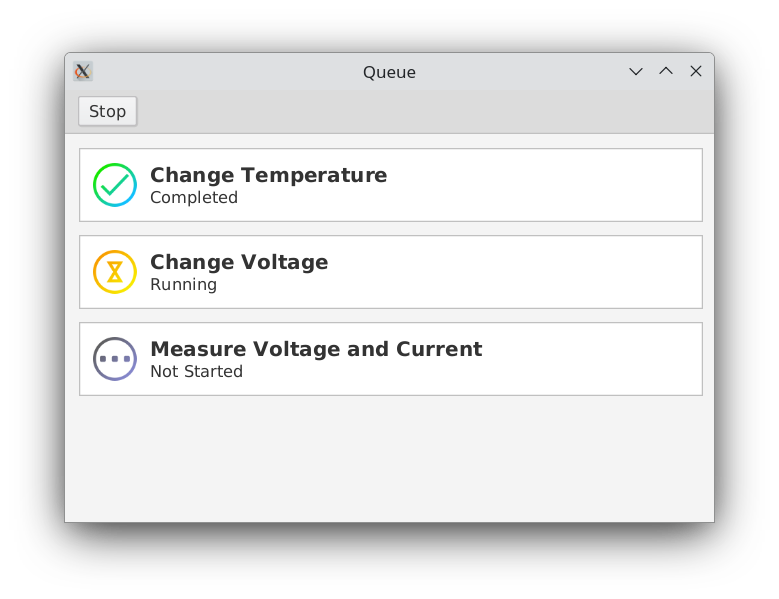}
\caption{An example of using the \mi{ActionQueue} to string together
separate actions, and the \mi{ActionQueueDisplay} GUI element to visually
present the current state of said queue to the user, with the resulting GUI
window when run shown below.\label{lst:queue}}
\end{listing*}
\newpage

\section{Impact on Research}

The JISA software library has enabled the relatively rapid automation of several
measurement systems within our laboratory, being used to drive many different
types of device measurement on field effect transistor architectures, including:
transistor output/transfer curves, electrical and thermal conductivity,
thermoelectric/Seebeck effect, Hall effect, and magnetoresistance. This
culminated in a combined piece of software called ``FetCh'' (from {\bf{F}ET}
{\bf{Ch}}aracterisation) which combines all these measurement types into a
single package, taking advantage of the queueing and measurement infrastructure
provided by the JISA library.

\begin{figure*}[!htbp]
    \centering
    \includegraphics[width=\linewidth]{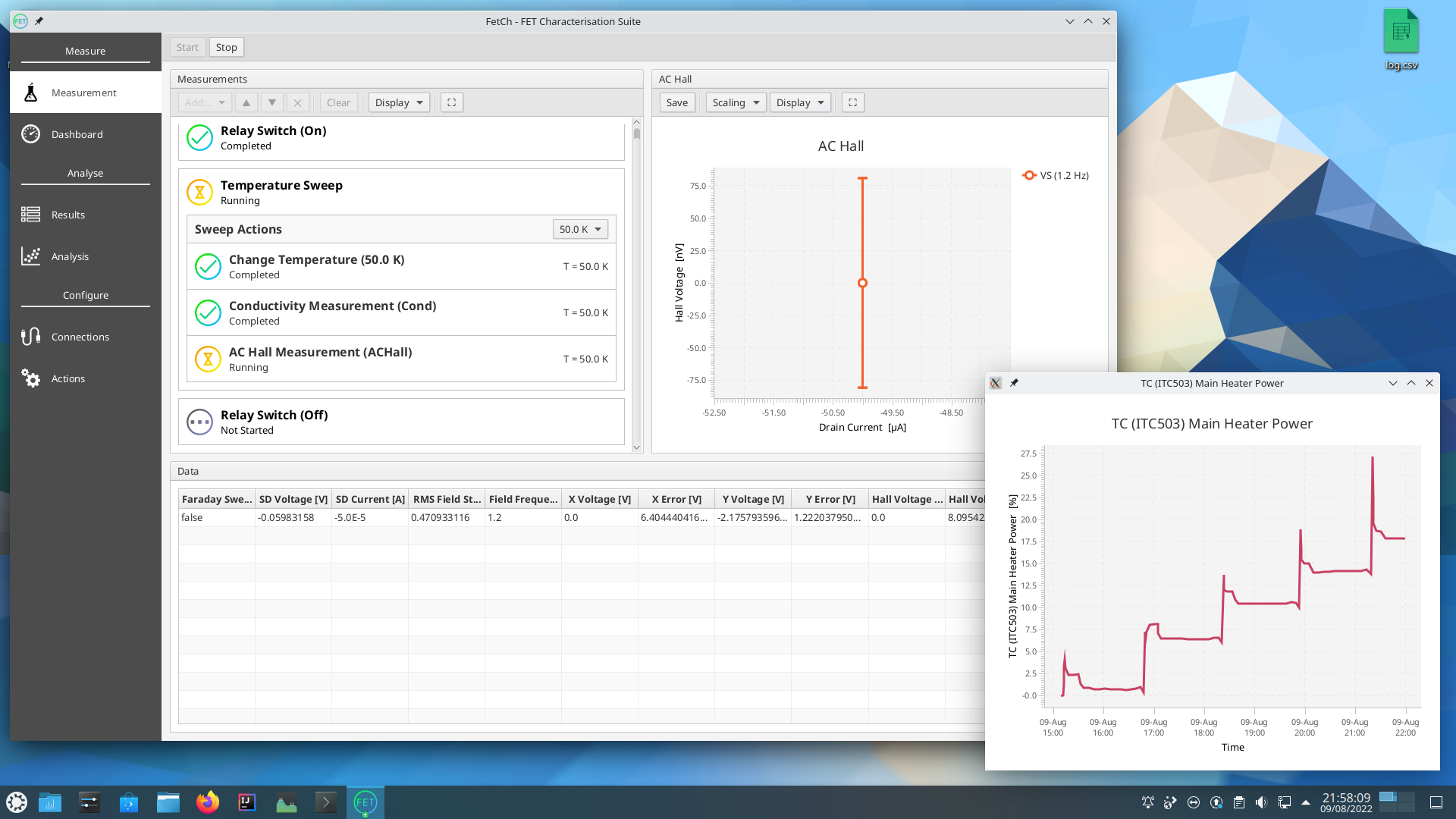}
    \caption{The FetCh software package, in use, and performing a temperature-dependent sweep of conductivity and Hall measurements.}
    \label{fig:fetch}
\end{figure*}

The underlying polymorphic nature of the JISA library then allows this single
piece of software to run on our different measurement systems without
modification, allowing its measurements to be run on any sufficiently featured
set of hardware. It simply requires that each piece of instrumentation used has
an appropriate driver class written for it (if it does not have one already).
The result of this is that many measurement systems in the lab gained a level of
automation which would previously have taken multiple different LabVIEW programs
or scripts to achieve. Furthermore, this means that users are only required to
learn how to use one piece of software to be able to control multiple systems.
This has resulted in several published works from the group that without FetCh,
and by extension the JISA library, would not have been possible to complete on
the same timescales.

Similarly, other pieces of software have been written for more specialised
systems, such as those for spectroscopy. AutoSpectra, written by Thomas Marsh,
is used for performing temperature-dependent, charge-modulated spectroscopy
measurements of various forms across multiple types of spectrometers, including
FTIR and UV-VIS. This software similarly takes advantage of the polymorphic
nature of the underlying JISA drivers, to enable it to work across these
multiple spectrometers, as well as temperature controllers (for temperature
dependence) and voltage sources (for charge modulation).

\subsection{List of enabled works}

\begin{itemize}

    \item Youcheng Zhang \emph{et al}. ``Direct Observation of Contact Reaction
    Induced Ion Migration and its Effect on Non-Ideal Charge Transport in Lead
    Triiodide Perovskite Field-Effect Transistors''. 2023. \cite{Zhang2023}
    
    \item William A. Wood \emph{et al}. ``Revealing contributions to conduction
    from transport within ordered and disordered regions in highly doped
    conjugated polymers through analysis of temperature-dependent Hall
    measurements''. FetCh used for AC Hall, and conductivity measurements. 2023.
    \cite{Wood2023}

    \item Dionisius H. L. Tjhe \emph{et al}. ``Thermoelectric Transport
    Signatures of Carrier Interactions in Polymer Electrochemical Transistors''.
    FetCh used for Seebeck, conductivity, and transistor measurements. 2023.
    \cite{Tjhe2023}

    \item Satyaprasad P. Senanayak \emph{et al}. ``Charge transport in mixed
    metal halide perovskite semiconductors''. FetCh used for AC Hall
    measurements. 2023. \cite{Senanayak2023}

    \item Ian E. Jacobs \emph{et al}. ``Structural and dynamic disorder, not
    ionic trapping, controls charge transport in highly doped conducting
    polymers''. FetCh used for AC Hall measurements. 2022. \cite{Jacobs2022}

    \item Yuxuan Huang \emph{et al}. ``Design of experiment optimization of
    aligned polymer thermoelectrics doped by ion-exchange''. FetCh used for
    Seebeck, and conductivity measurements. 2021. \cite{Huang2021}

    \item Martin Statz \emph{et al}. ``Charge and thermoelectric transport in
    polymer-sorted semiconducting single-walled carbon nanotube networks''.
    JISA-based software used for Seebeck, conductivity, and transistor
    measurements. 2020. \cite{Statz2020Charge}

    \item Martin Statz \emph{et al}, ``Temperature-Dependent Thermoelectric
    Transport in Polymer-Sorted Semiconducting Carbon Nanotube Networks with
    Different Diameter Distributions''. JISA-based software used for Seebeck,
    conductivity, and transistor measurements. 2020. \cite{Statz2020Temperature}

\end{itemize}

\section{Conclusions}

JISA is a software library created for standardising and building
test-and-measurement instrument control and automation software. By providing a
set of common interfaces of different instrument types, basic data handling
structures, and a suite of pre-defined GUI blocks, it allows one to rapidly
develop graphical test-and-measurement software in an instrument-agnostic way.
The main result being that a single measurement routine can be used on different
makes/models of instrument(s) without having to change the measurement routine
itself. Furthermore, it allows for this to be done without requiring extensive
programming knowledge or experience, and in many different programming languages
including: Java, Kotlin, and Python.

\newpage
\appendix
\section{Example Program (Output Curve)}

\begin{kotlincode}
fun outputCurve(drain: SMU, gate: SMU) {
    
    val SDV   = Column.ofDecimals("Source-Drain Voltage", "V")
    val SGV   = Column.ofDecimals("Source-Gate Voltage", "V")
    val DI    = Column.ofDecimals("Drain Current", "A")
    val GI    = Column.ofDecimals("Gate Current", "A")
    val data  = ResultList(SDV, SGV, DI, GI)
    
    val table = Table("Table of Results", data)
    val plot  = Plot("Output Curve")
    val grid  = Grid("Output Curve", table, plot)
    
    plot.createSeries().watch(data, SDV, DI).split(SGV)

    grid.addToolbarButton("Start") {

        data.clear()
        drain.setVoltage(0.0); gate.setVoltage(0.0);
        drain.turnOn(); gate.turnOn();

        for (g in Range.step(0, 50, 10)) {

            gate.setVoltage(g)

            for (d in Range.step(0, 50, 1)) {

                drain.setVoltage(d)
                
                data.addRow { row ->
                    row[SDV] = d
                    row[DI]  = drain.getCurrent()
                    row[SGV] = g
                    row[GI]  = gate.getCurrent()
                }

            }

        }

        drain.turnOff(); gate.turnOff();
    }

    grid.addToolbarButton("Save") {
        val file = GUI.saveFileSelect()
        if (file != null) {
            data.output(file)
        }
    }

}
\end{kotlincode}

\begin{figure}[htb]
    \centering
    \includegraphics[width=\linewidth]{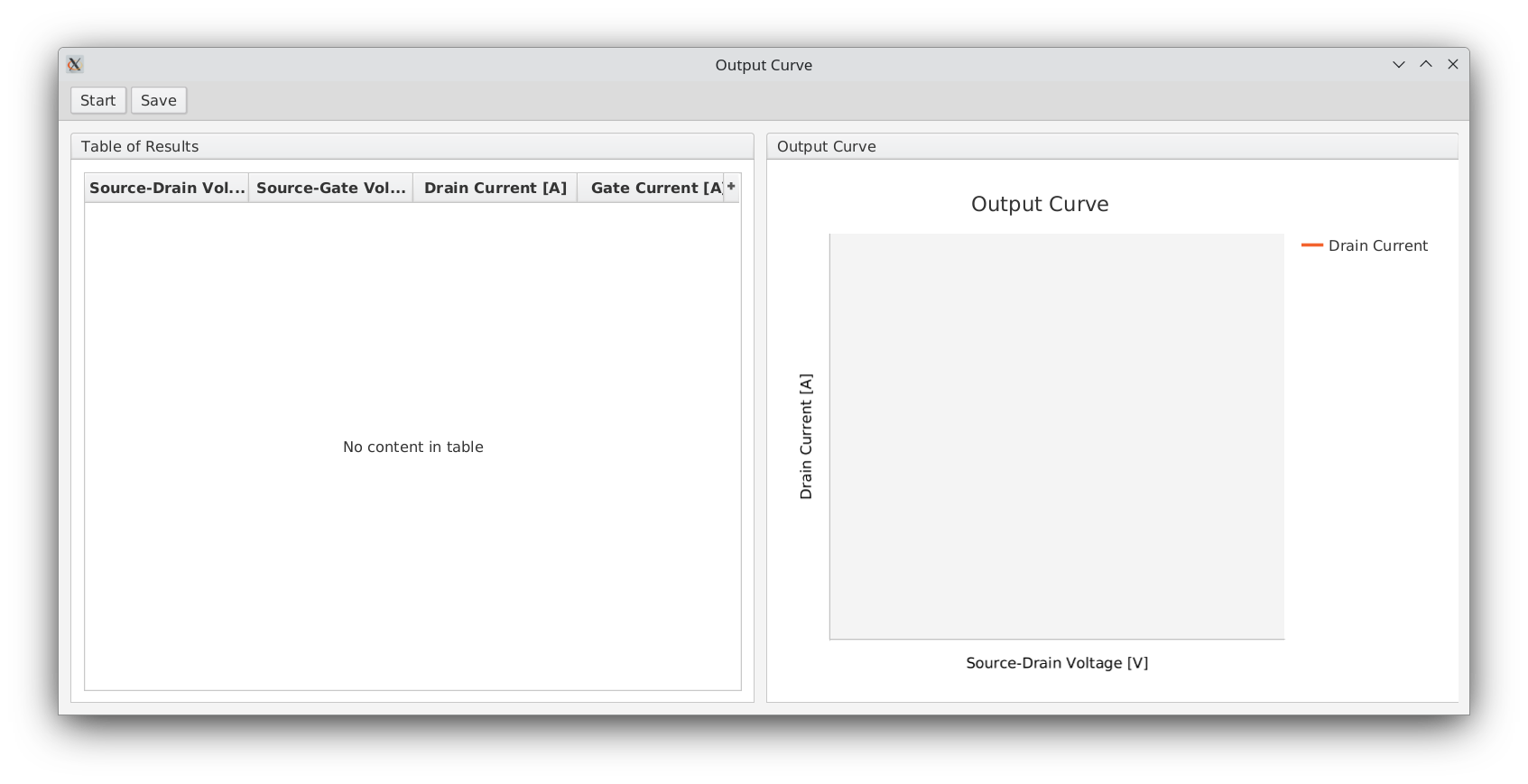}
    \includegraphics[width=\linewidth]{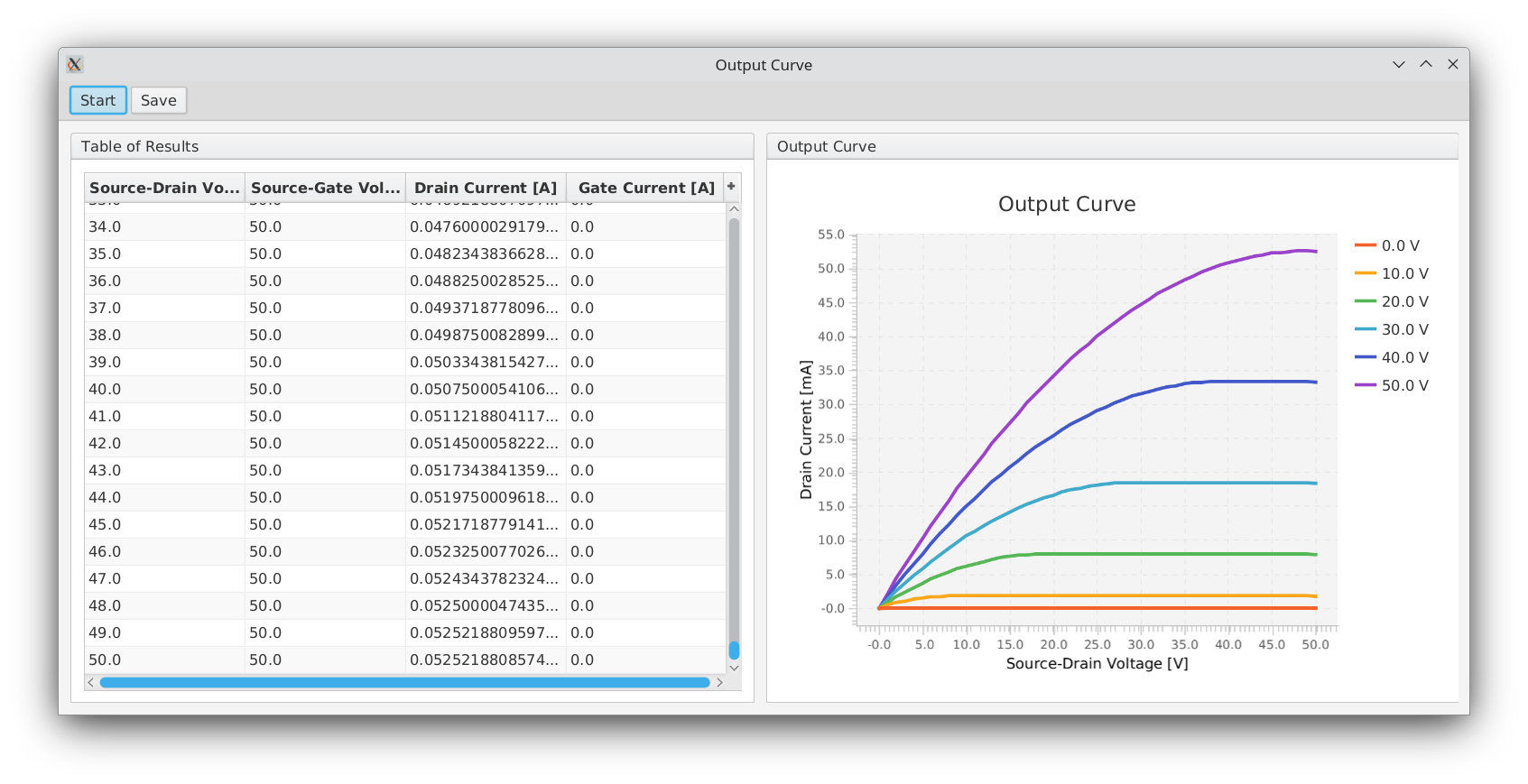}
    \caption{The result of running the example program (\mi{outputCurve(...)}),
    before pressing start (top), and after completing (bottom).\label{fig:output}}
\end{figure}

When this is run, it will produce the output as shown in
Figure~\ref{fig:output}. At first, both the table and plot will be blank/empty.
However, when the "Start" button is clicked, the output curve routine will run,
populating both. When the user clicks the "Save" button, they will be presented
with a save file dialogue, allowing them to specify where the output curve data
should be saved to in CSV format.

\bibliography{references}

\begin{thebibliography}{10}

\bibitem{IEEE488}
{IEEE Standard Digital Interface for Programmable Instrumentation}.
\newblock Standard, Institute of Electrical and Electronics Engineers, Nov
  1978.

\bibitem{IEEE4882}
{IEEE Standard Codes, Formats, Protocols, and Common Commands for Use With IEEE
  Std 488.1-1987, IEEE Standard Digital Interface for Programmable
  Instrumentation}.
\newblock Standard, Institute of Electrical and Electronics Engineers, Nov
  1992.

\bibitem{SCPI}
Standard commands for programmable instruments.
\newblock Standard, SCPI Consortium, IVI Foundation, Nov 1999.

\bibitem{IEEEDictionary}
{The Authoritative Dictionary of IEEE Standards Terms, Seventh Edition}.
\newblock {\em IEEE Std 100-2000}, page 574, 2000.

\bibitem{PyMeasure}
PyMeasure Team.
\newblock {PyMeasure}.
\newblock GitHub Repository \url{https://github.com/pymeasure/pymeasure}, 2022.

\bibitem{Mahmoud2004}
Qusay~H Mahmoud.
\newblock Practice and experience with java in education.
\newblock {\em Science of Computer Programming}, 53(1):1--2, 2004.

\bibitem{ChartFX}
FAIR~Facility for Anti-Proton and Ion Research.
\newblock {ChartFX}.
\newblock GitHub Repository \url{https://github.com/fair-acc/chart-fx}, 2022.

\bibitem{Zhang2023}
Youcheng Zhang, Amita Ummadisingu, Ravichandran Shivanna, Dionisius
  Hardjo~Lukito Tjhe, Hio-Ieng Un, Mingfei Xiao, Richard~H Friend,
  Satyaprasad~P Senanayak, and Henning Sirringhaus.
\newblock Direct observation of contact reaction induced ion migration and its
  effect on non-ideal charge transport in lead triiodide perovskite
  field-effect transistors.
\newblock {\em Small}, page 2302494, 2023.

\bibitem{Wood2023}
William~A Wood, Ian~E Jacobs, Leszek~J Spalek, Yuxuan Huang, Chen Chen,
  Xinglong Ren, and Henning Sirringhaus.
\newblock Revealing contributions to conduction from transport within ordered
  and disordered regions in highly doped conjugated polymers through analysis
  of temperature-dependent hall measurements.
\newblock {\em Physical Review Materials}, 7(3):034603, 2023.

\bibitem{Tjhe2023}
Dionisius Hardjo~Lukito Tjhe, Xinglong Ren, Ian Jacobs, Tarig Mustafa, Thomas
  Marsh, Yuxuan Huang, Lu~Zhang, William Wood, Ahmed Mansour, Gabriele d'Avino,
  et~al.
\newblock Thermoelectric transport signatures of carrier interactions in
  polymer electrochemical transistors.
\newblock {\em Bulletin of the American Physical Society}, 2023.

\bibitem{Senanayak2023}
Satyaprasad~P Senanayak, Krishanu Dey, Ravichandran Shivanna, Weiwei Li,
  Dibyajyoti Ghosh, Youcheng Zhang, Bart Roose, Szymon~J Zelewski, Zahra
  Andaji-Garmaroudi, William Wood, et~al.
\newblock Charge transport in mixed metal halide perovskite semiconductors.
\newblock {\em Nature Materials}, 22(2):216--224, 2023.

\bibitem{Jacobs2022}
Ian~E. Jacobs, Gabriele D’Avino, Vincent Lemaur, Yue Lin, Yuxuan Huang, Chen
  Chen, Thomas~F. Harrelson, William Wood, Leszek~J. Spalek, Tarig Mustafa,
  Christopher~A. O’Keefe, Xinglong Ren, Dimitrios Simatos, Dion Tjhe, Martin
  Statz, Joseph~W. Strzalka, Jin-Kyun Lee, Iain McCulloch, Simone Fratini,
  David Beljonne, and Henning Sirringhaus.
\newblock Structural and dynamic disorder, not ionic trapping, controls charge
  transport in highly doped conducting polymers.
\newblock {\em Journal of the American Chemical Society}, page jacs.1c10651, 2
  2022.

\bibitem{Huang2021}
Yuxuan Huang, Dionisius~Hardjo Lukito~Tjhe, Ian~E Jacobs, Xuechen Jiao, Qiao
  He, Martin Statz, Xinglong Ren, Xinyi Huang, Iain McCulloch, Martin Heeney,
  et~al.
\newblock Design of experiment optimization of aligned polymer thermoelectrics
  doped by ion-exchange.
\newblock {\em Applied Physics Letters}, 119(11), 2021.

\bibitem{Statz2020Charge}
Martin Statz, Severin Schneider, Felix~J Berger, Lianglun Lai, William~A Wood,
  Mojtaba Abdi-Jalebi, Simone Leingang, Hans-Jorg Himmel, Jana Zaumseil, and
  Henning Sirringhaus.
\newblock Charge and thermoelectric transport in polymer-sorted semiconducting
  single-walled carbon nanotube networks.
\newblock {\em ACS nano}, 14(11):15552--15565, 2020.

\bibitem{Statz2020Temperature}
Martin Statz, Severin Schneider, Felix Berger, Lianglun Lai, William Wood, Jana
  Zaumseil, and Henning Sirringhaus.
\newblock Temperature-dependent thermoelectric transport in polymer-sorted
  semiconducting carbon nanotube networks with different diameter
  distributions.
\newblock {\em Bulletin of the American Physical Society}, 65, 2020.

\end{thebibliography}

\end{document}